\documentclass[twocolumn,conference]{IEEEtran}
\usepackage{geometry}
\usepackage{setspace}
\usepackage[T1]{fontenc}
\usepackage[latin9]{inputenc}
\usepackage{verbatim}
\usepackage{float}
\usepackage{textcomp}
\usepackage{amsmath}
\usepackage{amsthm}
\usepackage{amssymb}
\usepackage{graphicx}
\usepackage{subfigure}
\makeatletter

\providecommand{\tabularnewline}{\\}
\floatstyle{ruled}
\newfloat{algorithm}{tbp}{loa}
\providecommand{\algorithmname}{Algorithm}
\floatname{algorithm}{\protect\algorithmname}
\usepackage{tikz}
\usetikzlibrary{calc}
\theoremstyle{plain}
\newtheorem{thm}{\protect\theoremname}
\theoremstyle{plain}
\newtheorem{lem}[thm]{\protect\lemmaname}

\usepackage{algorithmic}
\usepackage{cite}
\usepackage{siunitx}
\floatname{algorithm}{Algorithm}

\DeclareMathOperator{\maximize}{maximize}
\DeclareMathOperator{\minimize}{minimize}

\newcommand{\herm}{^{{\dagger}}}
\newcommand{\trans}{^{\mbox{\scriptsize T}}}
\providecommand{\lemmaname}{Lemma}
\providecommand{\theoremname}{Theorem}
\IEEEoverridecommandlockouts
\IEEEpubid{\makebox[\columnwidth]{978-1-7281-4490-0/20/\$31.00 \copyright~2020~IEEE }
	\hspace{\columnsep}\makebox[\columnwidth]{ }}
\IEEEaftertitletext{\vspace{-0.5\baselineskip}}

\begin{document}
\title{Accelerated Projected Gradient Method for the Optimization of Cell-Free
Massive MIMO Downlink}
\author{\IEEEauthorblockN{Muhammad~Farooq\IEEEauthorrefmark{1}, Hien~Quoc~Ngo\IEEEauthorrefmark{2},
and~Le~Nam~Tran\IEEEauthorrefmark{1}}\IEEEauthorblockA{\IEEEauthorrefmark{1}School of Electrical and Electronic Engineering,
University College Dublin, Ireland\\
Email: muhammad.farooq@ucdconnect.ie; nam.tran@ucd.ie}\IEEEauthorblockA{\IEEEauthorrefmark{2}Institute of Electronics, Communications and
Information Technology, Queen\textquoteright s University Belfast,
Belfast BT3 9DT, U.K.\\
Email: hien.ngo@qub.ac.uk}}
\maketitle
\begin{abstract}
We consider the downlink of a cell-free massive multiple-input multiple-output (MIMO) system where large number of access points (APs) simultaneously
serve a group of users. Two fundamental problems are of interest, namely (i) to maximize the total spectral efficiency (SE), and (ii) to maximize the minimum SE of all users. As the considered problems are non-convex, existing solutions rely
on successive convex approximation to find a sub-optimal solution. The known methods use off-the-shelf convex solvers, which basically implement an interior-point algorithm, to solve the derived convex problems. The main issue of such methods is that their complexity does not scale favorably with the problem size, limiting previous studies to cell-free massive MIMO of moderate scales. Thus the potential of cell-free massive MIMO has not been fully understood. To address
this issue, we propose an accelerated projected gradient method to solve the considered problems. Particularly, the proposed solution is found in closed-form expressions and only requires the first order information of the objective, rather than the Hessian matrix as in known solutions, and thus is much more memory efficient. Numerical results demonstrate that our proposed solution achieves far less run-time, compared to other second-order methods. 
\end{abstract}


\section{Introduction}

%Multiple-input multiple-output (MIMO) is the underlying technology
%in the physical layer of many modern wireless communications standards.
%The use of multiple antennas at transceivers can offer high data rates
%and high reliability by exploiting spatial and diversity gains. To
%meet a set of requirements for 5G networks, MIMO has evolved into
%so-called massive MIMO where a very large number of antennas are deployed
%at each base station (BS) \cite{Mazretta2010,Boccardi2014}. In particular,
%massive MIMO has been implemented in the first version of 5G NR \cite{Dahlman2018}. Since 5G still follows the conventional design of a \emph{cellular}
%network like its predecessors, inter-cell interference remains a fundamental
%problem, and thus massive MIMO cannot be unlocked to its full potential
%\cite{Lozano2013}.

Cell-free massive multiple-input multiple-output (MIMO) was introduced in \cite{Ngo2017} to overcome the inter-cell interference which is the main inherent limitation of cellular-based networks. In cell-free massive MIMO, many access points (APs) distributed
over the whole network serve many users in the same time-frequency
resource. There are no cells, and hence, no boundary effects. Unlike colocated massive MIMO, each AP is equipped with just a few antennas. But when the number of APs is very large, cell-free massive MIMO is still able to exploit the favorable propagation and channel hardening properties, like the colocated massive MIMO. In particular, in the downlink, each AP uses its local channel estimates acquired during
the uplink training phase to perform simple beamforming techniques. With this way, there is no need for exchanging the instantaneous channel
state information (CSI) among the APs or the central prcoessing unit (CPU) \cite{Ngo2018EE}.

The research on cell-free massive MIMO is still in its infancy and
thus deserves more extensive studies. We discuss here some of the
noticeable and related studies in the literature. In \cite{Ngo2017}, Ngo \emph{et al.}
considered the problem of minimum rate maximization  to provide uniformly good services to all users. The problem was then solved using a bisection search and a sequence of linear feasibility problems. In \cite{Nguyen2017}, Nguyen \emph{et al.} adopted zero-forcing precoding and studied the energy efficiency maximization (EEmax) problem. In this work, an iterative method based on successive convex approximation (SCA) was derived. In \cite{Buzzi2017}, both the max-min fairness and sum-rate maximization problems were considered and solved by SCA. The SCA-based method was also used in \cite{Ngo2018EE} and \cite{Interdonato2019} to solve  the EEmax and max-min fairness power controls with different cell-free massive MIMO setups, respectively.

A common feature of all the above mentioned pioneer studies on cell-free
massive MIMO is the use of  a second-order interior-point method which
requires the computation of the Hessian matrix of the objective, and thus  their computational complexity and memory requirement makes them infeasible in practical power control of large-scale cell-free massive MIMO.
It only allows us to characterize the performance of cell-free massive MIMO for a relatively small area.
For example, the work of \cite{Ngo2017} was able to consider an area of \SI{1}{\km}$\times$\SI{1}{\km},
consisting of $100$ APs serving $40$ users. Numbers with the same order of magnitude were also observed in the above mentioned papers.
The performance of these scenarios fractionally represents the full
potential of cell-free massive MIMO.

To fully understand the performance limits of cell-free massive MIMO, we need to devise more scalable resource allocation methods.
To this end, we propose in this paper a first order method to maximize the total spectral efficiency and the minimum rate of the downlink where the conjugate beamforming is adopted. 
A similar method has been used in \cite{Nam2019APG} to solve the EEmax problem.
%More precisely, 
In proposed method, we customize the accelerated proximal gradient (APG) method presented in \cite{Li2015}, which mainly requires the
computation of the gradient of the objective. 
The proposed method
is provably convergent and is numerically shown to achieve much lower
run-time compared to an existing second-order method given in \cite{Ngo2018EE}.

\emph{Notation}s: Bold
lower and upper case letters represent vectors and matrices. $\mathcal{CN}(0,a)$
denotes a complex Gaussian random variable with zero mean and variance
$a$. $\mathbf{X}\trans$ and $\mathbf{X}\herm$ stand for
the transpose and Hermitian of $\mathbf{X}$, respectively. $x_{i}$
is the $i$-th entry of vector $\mathbf{x}$; $[\mathbf{X}]_{i,j}$
is the $(i, j)$-th entry of $\mathbf{X}$.
$\nabla f(\mathbf{x})$ represents the gradient of $f(\mathbf{x})$
and $\frac{\partial}{\partial\mathbf{x}_{i}}f(\mathbf{x})$ is the
partial gradient with respect to $\mathbf{x}_{i}$. $[\mathbf{x}]_{+}$
denotes the projector onto the positive orthant. 
%When $\mathbf{X}_{1}$,
%..., $\mathbf{X}_{k}$ are matrices with the same number of columns,
%$[\mathbf{X}_{1};...;\mathbf{X}_{k}]$ stands for the matrix with
%the same number of columns obtained by staking vertically $\mathbf{X}_{1}$,
%..., and $\mathbf{X}_{k}$.
 $||\cdot||$ represents the Euclidean
norm; $|\cdot|$ is the absolute value of the argument. %

\section{System Model and Problem Formulation}
\subsection{System Model}
We consider the downlink of a cell-free massive MIMO system model as in \cite{Ngo2018EE}. In particular, there are $M$ APs serving $K$ single-antenna users in time division duplex (TDD) mode. Each AP is equipped with $N$ antennas. All the APs and the users are assumed to be distributed in a large area. As TDD operation
is adopted, APs first estimate the channels using pilot sequences from the uplink (commonly known as uplink training) and then apply a beamforming technique to transmit signals to all users in the downlink, or use a matched filter technique to detect signals in the uplink. Since this work focuses on the downlink transmission, we neglect the uplink phase. Let us denote by $T_{c}$
and $T_{p}$ the length of the coherence time and the uplink training phase in samples. The uplink training and downlink payload transmission phases are summarized as follows. The details can be found in \cite{Ngo2018EE}.

\subsubsection{Uplink Training and Downlink Payload Transmission}

We assume the channel is reciprocal, i.e., the channel gains on the uplink and on the downlink are the same. Consequently, APs can estimate
the downlink channel based on the pilot sequences sent by all users on the uplink. Let $\sqrt{T_{p}}\boldsymbol{\psi}_{k}\in\mathbb{C}^{T_{p}\times1}$,
where $||\boldsymbol{\psi}_{k}||^{2}=1$, be the pilot sequence transmitted
from the $k$-th user, $k=1,\ldots,K$.   As in \cite{Ngo2018EE}, we model the channel between the $m$-th AP and the $k$-th user as $$\mathbf{g}_{mk}=\beta_{mk}^{1/2}\mathbf{h}_{mk},$$
where $\text{\ensuremath{\beta_{mk}}}$ represents the large-scale
fading  and $\mathbf{h}_{mk}\in\mathbb{C}^{N\times1}$
comprises of small-scale fading coefficients between the $N$ antennas
of the $m$-th AP and the $k$-th user. We further assume that the
entries of $\mathbf{h}_{mk}$ follow i.i.d. $\mathcal{CN}(0,1)$.

Let $\mathbf{r}_{mk}=\mathbf{R}_{\textrm{up},m}\boldsymbol{\psi}_{k}$, where $\mathbf{R}_{\textrm{up},m}$
is the received signal at the $m$-th AP. Given $\mathbf{r}_{mk}$, the minimum mean-square
error (MMSE) estimate of ${\mathbf{g}}_{mk}$ is calculated as \cite{Ngo2018EE}
\begin{align}
\hat{\mathbf{g}}_{mk} & =\frac{\sqrt{\zeta_{p}T_{p}}\beta_{mk}}{1+\zeta_{p}T_{p}\sum_{i=1}^{K}\beta_{mi}\left|\boldsymbol{\psi}_{i}\herm\boldsymbol{\psi}_{k}\right|^{2}}\mathbf{r}_{mk},\label{eq:channel_estimate-1}
\end{align}
where $\zeta_{p}$ is the normalized transmit signal-to-noise ratio
(SNR) of each pilot symbol.  The mean square of any element of $\hat{\mathbf{g}}_{mk}$
is given by
\begin{equation}
\nu_{mk}=\mathbb{E}\{\left|[\hat{\mathbf{g}}_{mk}]_{n}\right|^{2}\}=\frac{\zeta_{p}T_{p}\beta_{mk}^{2}}{1+\zeta_{p}T_{p}\sum_{i=1}^{K}\beta_{mi}\left|\boldsymbol{\psi}_{i}\herm\boldsymbol{\psi}_{k}\right|^{2}}.
\end{equation}

For downlink payload data transmission, the APs use the channel estimates \eqref{eq:channel_estimate-1} and conjugate beamforming  to form separate radio
beams to the $K$ users. Denote the symbol to be sent to the $k$-th user by $c_{k}$ and the
power control coefficient between the $m$-th AP and the $k$-th user
by $\eta_{mk}$. Then the transmitted signal at the $m$-th AP  is 
\begin{align}
\mathbf{x}_{m}=\sqrt{\zeta_{d}}\sum\nolimits_{k=1}^{K}\sqrt{\eta_{mk}}\hat{\mathbf{g}}_{mk}^{*}c_{k},
\end{align}
where $\zeta_{d}$
is the maximum downlink transmit power at each AP normalized to the noise power.  Note that the total power at each AP is 
\begin{equation}
\mathbb{E}\{||\mathbf{x}_{m}||^{2}\}=\zeta_{d}N\sum\nolimits_{k=1}^{K}\eta_{mk}\nu_{mk}.\label{eq:spc:origin}
\end{equation}
The received signal at the $k$-th user is written
as 
\begin{align}
r_{k} & =\sum\nolimits_{m=1}^{M}\mathbf{g}_{mk}\trans\mathbf{x}_{m}+w_{k},\label{eq:receivedsig-1}
\end{align}
where $w_{k}\sim\mathcal{CN}(0,1)$ is the white Gaussian noise.

\subsubsection{Spectral Efficiency}
By using the use-and-then-forget capacity bounding technique, we can obtain the following spectral efficiency of the $k$-th user \cite{Ngo2018EE}
\begin{equation}
\mathrm{SE}_{k}(\bar{\boldsymbol{\eta}})=\Bigl(1-\frac{T_{p}}{T_{c}}\Bigr)\log_{2}\left(1+\gamma_{k}(\bar{\boldsymbol{\eta}})\right)\ (\textrm{bit/s/Hz}),\label{eq:SEk}
\end{equation}
where
\begin{equation}
\gamma_{k}(\bar{\boldsymbol{\eta}})=\frac{\zeta_{d}\left|\boldsymbol{\nu}_{kk}\trans\bar{\boldsymbol{\eta}}_{k}\right|^{2}}{\zeta_{d}\sum_{i\neq k}^{K}\left|\boldsymbol{\nu}_{ik}\trans\bar{\boldsymbol{\eta}}_{i}\right|^{2}+\frac{\zeta_{d}}{N}\sum_{i=1}^{K}||\mathbf{D}_{ik}\bar{\boldsymbol{\eta}}_{i}||_{2}^{2}+\frac{1}{N^2}}
\end{equation}
and where $\bar{\boldsymbol{\eta}}_{k}=[\sqrt{\eta_{1k}};\ldots;\sqrt{\eta_{Mk}}]\in\mathbb{R}_{+}^{M}$
consists of all power control coefficients associated with user $k$,
$\bar{\boldsymbol{\eta}}=[\bar{\boldsymbol{\eta}}_{1};\bar{\boldsymbol{\eta}}_{2};\ldots;\bar{\boldsymbol{\eta}}_{K}]\in\mathbb{R}_{+}^{MK}$,
$\mathbf{D}_{ik}\in\mathbb{R}_{+}^{M\times M}$ is a diagonal matrix
with $[\mathbf{D}_{ik}]_{m,m}=\sqrt{\nu_{mi}\beta_{mk}}$, and $\boldsymbol{\nu}_{ik}\triangleq\left|\boldsymbol{\psi}_{i}\herm\boldsymbol{\psi}_{k}\right|\left[\nu_{1i}\frac{\beta_{1k}}{\beta_{1i}};\nu_{2i}\frac{\beta_{2k}}{\beta_{2i}};\ldots;\nu_{Mi}\frac{\beta_{Mk}}{\beta_{Mi}}\right].$

\subsection{Problem Formulation}

To formulate the considered problem and to facilitate the development
of the proposed algorithm, we define $\boldsymbol{\mu}_{m}\in\mathbb{R}_{+}^{K}$
to be the vector of all power control coefficients associated with
the $m$-th AP as $\boldsymbol{\mu}_{m}\triangleq[\mu_{m1};\mu_{m2};\ldots;\mu_{mK}]$,
where $\mu_{mk}=\sqrt{\eta_{mk}\nu_{mk}}, m=1,\ldots,M, k=1,\ldots,K.$
We also define $\boldsymbol{\mu}\triangleq[\boldsymbol{\mu}_{1};\boldsymbol{\mu}_{2};\ldots;\boldsymbol{\mu}_{M}]\in\mathbb{R}_{+}^{MK\times1}$
to include the power control coefficients of all APs. To express the
spectral efficiency in \eqref{eq:SEk} as a function of $\boldsymbol{\mu}$,
we denote by $\bar{\boldsymbol{\mu}}_{k}=[\mu_{1k};\mu_{2k};\ldots;\mu_{Mk}]$
the vector of power control coefficients associated with user $k$.
Thus we can write $\boldsymbol{\nu}_{ik}\trans\boldsymbol{\eta}_{i}$
as $\bar{\boldsymbol{\nu}}_{ik}\bar{\boldsymbol{\mu}}_{k},$ where
%\begin{equation}
$\bar{\boldsymbol{\nu}}_{ik}\triangleq\Bigl|\boldsymbol{\psi}_{i}\herm\boldsymbol{\psi}_{k}\Bigr|\Bigl[\sqrt{\nu_{1i}}\frac{\beta_{1k}}{\beta_{1i}};\sqrt{\nu_{2i}}\frac{\beta_{2k}}{\beta_{2i}};\ldots;\sqrt{\nu_{Mi}}\frac{\beta_{Mk}}{\beta_{Mi}}\Bigr].$
%\end{equation}
Similarly, we can write $\mathbf{D}_{ik}\bar{\boldsymbol{\eta}}_{i}$
as $\bar{\mathbf{D}}_{i}\bar{\boldsymbol{\mu}}_{i}$, where $\bar{\mathbf{D}}_{i}$
is a diagonal matrix with the $m$-th diagonal entry equal to $\sqrt{\beta_{mi}}$.
Now the spectral efficiency of the $k$-th user (in nat/s/Hz) can
be expressed as
\begin{equation}
\mathrm{SE}_{k}(\boldsymbol{\mu})=\bigl(1-\frac{T_{p}}{T_{c}}\bigr)\log\left(1+\gamma_{k}(\boldsymbol{\mu})\right),\label{eq:SEk_mu}
\end{equation}
where $\gamma_{k}(\boldsymbol{\mu})$ is the SINR of the $k$-th user
given by
\begin{equation}
\gamma_{k}(\boldsymbol{\mu})=\frac{\zeta_{d}(\bar{\boldsymbol{\nu}}_{kk}\trans\bar{\boldsymbol{\mu}}_{k})^{2}}{\zeta_{d}\sum_{i\neq k}^{K}(\bar{\boldsymbol{\nu}}_{ik}\trans\bar{\boldsymbol{\mu}}_{i})^{2}+\frac{\zeta_{d}}{N}\sum_{i=1}^{K}||\bar{\mathbf{D}}_{i}\bar{\boldsymbol{\mu}}_{i}||_{2}^{2}+\frac{1}{N^2}}.
\end{equation}
%\newpage
The total spectral efficiency of the system is defined as 
\begin{equation}
\mathrm{SE}(\boldsymbol{\mu})\triangleq\sum\nolimits _{k=1}^{K}\mathrm{SE}_{k}(\boldsymbol{\mu}).\label{eq:SEsum}
\end{equation}
In this paper, we consider a total power constraint at each AP which is given by $\mathbb{E}\{||\mathbf{x}_{m}||^{2}\}\leq\zeta_{d}$ or $||\boldsymbol{\mu}_{m}||^{2}\leq\frac{1}{N}, m=1,2,\ldots,M$,
which follows from \eqref{eq:spc:origin}. For the problem formulation
purpose, we define the following set $\mathcal{S}=\left\{ \boldsymbol{\mu}|\boldsymbol{\mu}\geq0;||\boldsymbol{\mu}_{m}||^{2}\leq 1/N, m=1,2,\ldots,M\right\} .$
In this paper, we consider the following two common power control optimization problems,
namely
\begin{itemize}
\item The problem of maximizing the total spectral efficiency (SEmax)
\begin{equation}
\boxed{(\mathcal{P}_{1}):\underset{\boldsymbol{\mu}}{\maximize}\ \Bigl\{\sum\nolimits _{k=1}^{K}\mathrm{SE}_{k}(\boldsymbol{\mu})\ |\ \boldsymbol{\mu}\in\mathcal{S}\Bigr\}}\label{eq:SEmaxproblem}
\end{equation}
\item The problem of maximizing the minimum rate (MRmax) among all users (also known
as max-min fairness maximization) 
\begin{equation}
\boxed{(\mathcal{P}_{2}):\underset{\boldsymbol{\mu}}{\maximize}\ \Bigl\{\min_{1\leq k\leq K}\mathrm{SE}_{k}(\boldsymbol{\mu})\ |\ \boldsymbol{\mu}\in\mathcal{S}\Bigr\}}\label{eq:minrate}
\end{equation}
\end{itemize}
For the above problems, SCA has 
%proved to be very effective and 
gradually
become a standard mathematical tool \cite{Ngo2017,Ngo2018EE}. 
%The
%idea of SCA is to approximate a non-convex program by a series of
%convex sub-problems. 
In all known solutions for the considered problems
or related ones, interior point methods (through the use of off-the-shelf
convex solvers) are invoked to solve these convex problems \cite{Ngo2018EE,Buzzi2017,Interdonato2019},
which do not scale with the problem size. In the next section, we propose
methods that can tackle this scalability problem.

\section{Proposed Solutions}

In this section, we present solutions to $(\mathcal{P}_{1})$ and
$(\mathcal{P}_{2})$, using the accelerated proximal gradient (APG)
methods introduced in \cite{Li2015}, 
%In general the two considered
%problems can be written in a compact form as \begin{subequations}\label{eq:modifiedproblem}
%\begin{align}
%\underset{\boldsymbol{\mu}}{\maximize} & \quad f(\boldsymbol{\mu})\\
%\st & \quad\boldsymbol{\mu}\in\mathcal{S},
%\end{align}
%\end{subequations} where $f(\boldsymbol{\mu})=\sum\nolimits _{k=1}^{K}\mathrm{SE}_{k}(\boldsymbol{\mu})$
%for $(\mathcal{P}_{1})$ and $f(\boldsymbol{\mu})=\min_{1\leq k\leq K}\mathrm{SE}_{k}(\boldsymbol{\mu})$
%for $(\mathcal{P}_{2})$. 
%We remark that the methods described in
%\cite{Li2015} 
which concerns the following problem
\begin{equation}
\underset{\boldsymbol{\mu}\in\mathbb{R}^{n}}{\min}\ \{F(\boldsymbol{\mu})\equiv f(\boldsymbol{\mu})+g(\boldsymbol{\mu})\},\label{eq:nonconvex:gen}
\end{equation}
where $f(\boldsymbol{\mu})$ is differentiable (but possibly \emph{nonconvex})
and $g(\mathbf{x})$ can be both nonconvex and \emph{nonsmooth}. If
we let $g(\boldsymbol{\mu})$ be the indicator function of $\mathcal{S}$,
defined as
\begin{equation}
\delta_{\mathcal{S}}(\boldsymbol{\mu})=\begin{cases}
0 & \boldsymbol{\mu}\in\mathcal{S}\\
+\infty & \boldsymbol{\mu}\notin\mathcal{S},
\end{cases}
\end{equation}
then \eqref{eq:nonconvex:gen} is actually equivalent to $(\mathcal{P}_{1})$. Basically the objective of $(\mathcal{P}_{2})$ is nonsmooth, but we can still solve $(\mathcal{P}_{2})$ by applying a proper smoothing technique. 
We also note that when $g(\boldsymbol{\mu})$ is the indicator function of $\mathcal{S}$, the proximal operator of $g(\boldsymbol{\mu})$ becomes the
Euclidean projection onto $\mathcal{S}$. In the following, we customize
the APG methods to solve the considered problems.

\subsection{Proposed Solution to $(\mathcal{P}_{1})$}

Since $f(\boldsymbol{\mu})$ for $(\mathcal{P}_{1})$ is differentiable,
the proposed algorithm for solving $(\mathcal{P}_{1})$ follows closely
the monotone APG method in \cite{Li2015}, which is outlined in Algorithm
\ref{alg:mAPG}. In Algorithm \ref{alg:mAPG}, $\alpha>0$ is called
the step size which should be sufficiently small to guarantee its
convergence. Also, the notation $P_{\mathcal{S}}(\mathbf{u})$ denotes
the projection onto $\mathcal{S}$, i.e., $P_{\mathcal{S}}(\mathbf{u})=\arg\min\bigl\{||\mathbf{x}-\mathbf{u}||\ |\ \mathbf{x}\in\mathcal{S}\bigr\}$.
From a given operating point, we move along the direction of the gradient
of $f(\boldsymbol{\mu})$ with the step size $\alpha$, and then project
the resulting point onto the feasible set. In particular, $\mathbf{y}^{n}$
in Step \ref{alg:mAPG:extra} is an extrapolated point which is used
for convergence acceleration. However, $\mathbf{y}^{n}$ can be a bad extrapolation unlike APG methods for the
convex case, and thus
Step \ref{alg:mAPG:update} is there to fix this issue.%
\begin{algorithm}[th]
	\small
\caption{General Description of Proposed Algorithm for Solving $(\mathcal{P}_{1})$
and $(\mathcal{P}_{2})$}
\label{alg:mAPG}

\begin{algorithmic}[1]

\STATE Input: $\boldsymbol{\mu}^{0}>=0,t_{0}=t_{1}=1,\alpha>0$

\STATE $\boldsymbol{\mu}^{1}=\mathbf{z}^{1}=\boldsymbol{\mu}^{0}$

\FOR{ $n=1,2,\ldots$}

\STATE $\mathbf{y}^{n}=\boldsymbol{\mu}^{n}+\frac{t_{n-1}}{t_{n}}(\mathbf{z}^{n}-\boldsymbol{\mu}^{n})+\frac{t_{n-1}-1}{t_{n}}(\boldsymbol{\mu}^{n}-\boldsymbol{\mu}^{n-1})$\label{alg:mAPG:extra}

\STATE $\mathbf{z}^{n+1}=P_{\mathcal{S}}(\mathbf{y}^{n}+\alpha\nabla f(\mathbf{y}^{n}))$
\label{grady}

\STATE $\mathbf{v}^{n+1}=P_{\mathcal{S}}(\boldsymbol{\mu}^{n}+\alpha\nabla f(\boldsymbol{\mu}^{n}))$\label{gradmu}

\STATE $\boldsymbol{\mu}^{n+1}=\begin{cases}
\mathbf{z}^{n+1} & f(\mathbf{z}^{n+1})\geq f(\mathbf{v}^{n+1})\\
\mathbf{v}^{n+1} & \textrm{otherwise}
\end{cases}$\label{alg:mAPG:update}

\STATE $t_{n+1}=0.5\left(\sqrt{4t_{n}^{2}+1}+1\right)$

\ENDFOR

\STATE Output: $\boldsymbol{\mu}^{*}$

\end{algorithmic}
\end{algorithm}
 We now give the details for the two main operations of Algorithm
\ref{alg:mAPG}, namely: the projection onto the feasible set $\mathcal{S}$
and the gradient of $f(\boldsymbol{\mu})$.

\subsubsection{Projection onto $\mathcal{S}$}

We show that the projection in Steps \ref{grady} and \ref{gradmu}
in Algorithm \ref{alg:mAPG} can be done \emph{in parallel} and by
\emph{closed-form expressions}. Recall that for a given  $\mathbf{x}\in\mathbb{R}^{MK\times1}$,
$P_{\mathcal{S}}(\mathbf{x})$ is the solution to the following problem
\begin{subequations}
\begin{multline}
\underset{\boldsymbol{\mu}\in\mathbb{R}^{MK\times1}}{\minimize}\ \Bigl\{||\boldsymbol{\mu}-\mathbf{x}||^{2}\ \Bigl|\ \boldsymbol{\mu}\geq0;||\boldsymbol{\mu}_{m}||^{2}\leq\frac{1}{N},\\
m=1,2,\ldots,M\Bigr\}.
\end{multline}
\end{subequations} It is easy to see that the above problem can
be decomposed into sub-problems at each AP $m$ as \begin{subequations}\label{eq:project:subprob}
\begin{align}
\underset{\boldsymbol{\mu}_{m}\in\mathbb{R}^{K\times1}}{\minimize} & \Bigl\{||\boldsymbol{\mu}_{m}-\mathbf{x}_{m}||^{2}\ \Bigl|\ \boldsymbol{\mu}_{m}\geq0;||\boldsymbol{\mu}_{m}||^{2}\leq\frac{1}{N}\Bigr\}.
\end{align}
\end{subequations}The above problem is in fact the projection onto
the intersection of a ball and the positive orthant. Interestingly,
the analytical solution to this problem can be found by applying \cite[Theorem 7.1]{Bauschke2017},
which produces
\begin{equation}
\boldsymbol{\mu}_{m}=\frac{\sqrt{1/N}}{\max\left(\sqrt{1/N},||[\mathbf{x}_{m}]_{+}||\right)}[\mathbf{x}_{m}]_{+}.\label{eq:projectionEuclidean}
\end{equation}
%The above expression means that we first project $\mathbf{x}_{m}$
%onto the positive orthant and then onto the Euclidean ball of radius
%$\sqrt{1/N}$. 
\subsubsection{Gradient of $f(\boldsymbol{\mu})$ for $(\mathcal{P}_{1})$}
To implement Algorithm \ref{alg:mAPG}, we also need to compute $\nabla_{\boldsymbol{\mu}}f(\boldsymbol{\mu})$,
%which is derived in what follows. We know that the gradient of a multi-variable
%function is the vector of all its partial derivatives, i.e.
which is found as\begin{align}
\nabla f(\boldsymbol{\mu}) & =[\frac{\partial}{\partial\bar{\boldsymbol{\mu}}_{1}}f(\boldsymbol{\mu});\frac{\partial}{\partial\bar{\boldsymbol{\mu}}_{2}}f(\boldsymbol{\mu}),\ldots,\frac{\partial}{\partial\bar{\boldsymbol{\mu}}_{K}}f(\boldsymbol{\mu})],\label{eq:gradfmu}
\end{align}
where $\frac{\partial}{\partial\bar{\boldsymbol{\mu}}_{i}}f(\boldsymbol{\mu})=\sum_{k=1}^{K}\frac{\partial}{\partial\bar{\boldsymbol{\mu}}_{i}}\mathrm{SE}_{k}(\boldsymbol{\mu})$.
Thus it basically boils down to finding $\frac{\partial}{\partial\bar{\boldsymbol{\mu}}_{i}}\mathrm{SE}_{k}(\boldsymbol{\mu})$.
To this end, let us define $b_{k}(\boldsymbol{\mu})=\zeta_{d}(\bar{\boldsymbol{\nu}}_{kk}\trans\bar{\boldsymbol{\mu}}_{k})^{2}$
and $c_{k}(\boldsymbol{\mu})=\zeta_{d}\left(\sum_{i\neq k}^{K}(\bar{\boldsymbol{\nu}}_{ik}\trans\bar{\boldsymbol{\mu}}_{i})^{2}+\frac{1}{N}\sum_{i=1}^{K}||\bar{\mathbf{D}}_{i}\bar{\boldsymbol{\mu}}_{i}||_{2}^{2}\right)+\frac{1}{N^{2}}$.
Then we can rewrite $\mathrm{SE}_{k}(\boldsymbol{\mu})$ \eqref{eq:SEk}
as
\begin{equation}
\mathrm{SE}_{k}(\boldsymbol{\mu})=\log\bigl(b_{k}(\boldsymbol{\mu})+c_{k}(\boldsymbol{\mu})\bigr)-\log c_{k}(\boldsymbol{\mu}).\label{eq:SElogterm}
\end{equation}
The gradient of $\mathrm{SE}_{k}(\boldsymbol{\mu})$ with respect
to $\bar{\boldsymbol{\mu}}_{i}$, $i=1,2,\ldots,K$, is found as 
\begin{equation}
\frac{\partial}{\partial\bar{\boldsymbol{\mu}}_{i}}\mathrm{SE}_{k}(\boldsymbol{\mu})=\frac{\frac{\partial}{\partial\bar{\boldsymbol{\mu}}_{i}}\left(b_{k}(\boldsymbol{\mu})+c_{k}(\boldsymbol{\mu})\right)}{b_{k}(\boldsymbol{\mu})+c_{k}(\boldsymbol{\mu})}-\frac{\frac{\partial}{\partial\bar{\boldsymbol{\mu}}_{i}}c_{k}(\boldsymbol{\mu})}{c_{k}(\boldsymbol{\mu})}.\label{eq:gradGamma}
\end{equation}
Now we recall the following equality $\nabla||\mathbf{A}\mathbf{x}||^{2}=2\mathbf{A}\trans\mathbf{A}\mathbf{x}$
for any symmetric matrix $\mathbf{A}$, and thus $\nabla_{\bar{\boldsymbol{\mu}}_{i}}b_{k}(\boldsymbol{\mu})$
and $\nabla_{\bar{\boldsymbol{\mu}}_{i}}c_{k}(\boldsymbol{\mu})$
are respectively given by 
\begin{equation}
\frac{\partial}{\partial\bar{\boldsymbol{\mu}}_{i}}b_{k}(\boldsymbol{\mu})=\begin{cases}
2\zeta_{d}\bar{\boldsymbol{\nu}}_{kk}\bar{\boldsymbol{\nu}}_{kk}\trans\bar{\boldsymbol{\mu}}_{k}, & i=k\\
0, & i\neq k
\end{cases}\label{eq:BkGradient}
\end{equation}

\begin{equation}
\frac{\partial}{\partial\bar{\boldsymbol{\mu}}_{i}}c_{k}(\boldsymbol{\mu})=\begin{cases}
2(\zeta_{d}/N)\mathbf{\bar{D}}_{k}^{2}\bar{\boldsymbol{\mu}}_{k}, & i=k\\
2\zeta_{d}\bar{\boldsymbol{\nu}}_{ik}\bar{\boldsymbol{\nu}}_{ik}\trans\bar{\boldsymbol{\mu}}_{i}+\frac{2\zeta_{d}}{N}\mathbf{\bar{D}}_{i}^{2}\bar{\boldsymbol{\mu}}_{i}, & i\neq k
\end{cases}.\label{eq:CkGradient}
\end{equation}

\subsection{Proposed Solution to $(\mathcal{P}_{2})$}

We recall that for $(\mathcal{P}_{2})$ the objective function is
\begin{equation}
f(\boldsymbol{\mu})=\underset{1\leq k\leq K}{\min}\mathrm{SE}_{k}(\boldsymbol{\mu}),
\end{equation}
which is \emph{non-differentiable}. Thus a straightforward application
of the APG method is impossible. To overcome this issue, we adopt
a smoothing technique. In particular, $f(\boldsymbol{\mu})$ is approximated
by the following log-sum-exp function given by\cite{Nesterov2005a}
\begin{equation}
f_{\tau}(\boldsymbol{\mu})=-\frac{1}{\tau}\log\Bigl(\frac{1}{K}\sum\nolimits _{k=1}^{K}\exp\bigl(-\tau\mathrm{SE}_{k}(\boldsymbol{\mu})\Bigr),\label{eq:smoothapprox}
\end{equation}
where $\tau>0$ is the positive \emph{smoothness} parameter. To obtain
\eqref{eq:smoothapprox}, we have used the fact that $\underset{1\leq k\leq K}{\min}\mathrm{SE}_{k}(\boldsymbol{\mu})=-\underset{1\leq k\leq K}{\max}-\mathrm{SE}_{k}(\boldsymbol{\mu})$.
Nesterov proved in \cite{Nesterov2005a} that $f(\boldsymbol{\mu})+\frac{\log K}{\tau}\geq f_{\tau}(\boldsymbol{\mu})\geq f(\boldsymbol{\mu})$.
In other words, $f_{\tau}(\boldsymbol{\mu})$ is a differentiable
approximation of $f(\boldsymbol{\mu})$ with a numerical accuracy
of $\frac{\log K}{\tau}$. Thus, with a sufficiently high $\tau$,
we can find an approximate solution to $(\mathcal{P}_{2})$ by running
Algorithm \ref{alg:mAPG} with $f(\boldsymbol{\mu})$ being replaced
by $f_{\tau}(\boldsymbol{\mu})$ in \eqref{eq:smoothapprox}. In this
regard, the gradient of $f_{\tau}(\boldsymbol{\mu})$ is easily found
as $\nabla_{\bar{\boldsymbol{\mu}}}f_{\tau}(\boldsymbol{\mu})=[\frac{\partial}{\partial\bar{\boldsymbol{\mu}}_{1}}f_{\tau}(\boldsymbol{\mu}),\frac{\partial}{\partial\bar{\boldsymbol{\mu}}_{2}}f_{\tau}(\boldsymbol{\mu}),\ldots,\frac{\partial}{\partial\bar{\boldsymbol{\mu}}_{K}}f_{\tau}(\boldsymbol{\mu})]$,
where $\frac{\partial}{\partial\bar{\boldsymbol{\mu}}_{i}}f_{\tau}(\boldsymbol{\mu})$
is given by 
\begin{equation}
\frac{\partial}{\partial\bar{\boldsymbol{\mu}}_{i}}f_{\tau}(\boldsymbol{\mu})=\frac{\sum_{k=1}^{K}\Bigl(\exp\bigl(-\tau\mathrm{SE}_{k}(\boldsymbol{\mu})\bigr)\nabla_{\bar{\boldsymbol{\mu}}_{k}}\mathrm{SE}_{k}(\boldsymbol{\mu})\Bigr)}{\sum_{k=1}^{K}\exp\bigl(-\tau\mathrm{SE}_{k}(\boldsymbol{\mu})\bigr)}.
\end{equation}

\section{Complexity and Convergence Analysis of Proposed Methods}

\subsection{Complexity Analysis}

We now provide the complexity analysis of the proposed algorithm
for one iteration using the big-O notation. It is clear that the complexity
of Algorithm \ref{alg:mAPG} is dominated by the computation of
three quantities: the objective, the gradient, and the projection.
It is easy to see that $KM$ multiplications are required to compute
$\mathrm{SE}_{k}(\boldsymbol{\mu})$. Therefore, the complexity of
finding $f(\boldsymbol{\mu})=\sum_{k=1}^{K}\mathrm{SE}_{k}(\boldsymbol{\mu})$
is $\mathcal{O}(K^{2}M).$ Similarly, we can find that the
complexity of $\nabla\mathrm{SE}(\boldsymbol{\mu})$ which is $\mathcal{O}(K^{2}M)$.
The projection of $\boldsymbol{\mu}$ onto $\mathcal{S}$ is given
in \eqref{eq:projectionEuclidean}, which requires the computation
of the $l_{2}$-norm of $K\times1$ vector $\mathbf{x}_{m}$ at each
AP, and thus the complexity of the projection is $\mathcal{O}(KM)$.
In summary, the per-iteration complexity of the proposed algorithm
for solving $(\mathcal{P}_{1})$ is $\mathcal{O}(K^{2}M)$. We can
also conclude that the per-iteration complexity for solving $(\mathcal{P}_{2})$
is also $\mathcal{O}(K^{2}M)$.

\subsection{Convergence Analysis}

We now discuss the convergence result of Algorithm \ref{alg:mAPG}
for solving $(\mathcal{P}_{1})$ and $(\mathcal{P}_{2})$. For $(\mathcal{P}_{1})$,
from \eqref{eq:gradGamma}, \eqref{eq:BkGradient}, \eqref{eq:CkGradient},
we can check that $\nabla f(\boldsymbol{\mu})$ is Lipschitz continuous,
or equivalently $f(\boldsymbol{\mu})$ has Lipschitz continuous gradient.
That is, there exists a constant $L>0$ such that 
\begin{equation}
||\nabla f(\mathbf{x})-\nabla f(\mathbf{\mathbf{y}})||\leq L||\mathbf{x}-\mathbf{y}||\label{eq:gradLipschitz-1}.
\end{equation}
 For the considered problems, it is possible to find a Lipschitz
constant for $\nabla f(\boldsymbol{\mu})$ but we skip the details
here for the sake of brevity. Let us simply denote by $L$ the Lipschitz
constant of $\nabla f(\boldsymbol{\mu})$. The convergence of Algorithm
\ref{alg:mAPG} is stated in the following lemma.
\begin{lem}
Let the step size $\alpha$ in Algorithm\textup{ \ref{alg:mAPG}}
satisfy $\alpha<\frac{1}{L}$. Then the iterates $\{\boldsymbol{\mu}^{n}\}$
generated by Algorithm \ref{alg:mAPG} are bounded. Let $\boldsymbol{\mu}^{\ast}$
be any accumulation point of $\{\boldsymbol{\mu}^{n}\}$, then $\boldsymbol{\mu}^{\ast}$
is a critical point of $(\mathcal{P}_{1})$.
\end{lem}
\begin{IEEEproof}
Please see Appendix \ref{sec:Convergence-Proof}.
\end{IEEEproof}
The same convergence result applies to Algorithm \ref{alg:mAPG} for
solving $(\mathcal{P}_{2})$ with the differential approximation in
\eqref{eq:smoothapprox}.

\subsection{Improved Convergence with Line Search}

In practice, we do not need to compute a Lipschitz constant
of $\nabla f(\boldsymbol{\mu})$ for two reasons. First, the best
Lipschitz constant of $\nabla f(\boldsymbol{\mu})$ (i.e. the smallest
$L$ such that \eqref{eq:gradLipschitz-1} holds) is hard to find.
Second, the conditions $\alpha<\frac{1}{L}$ is sufficient but not
necessary for Algorithm \ref{alg:mAPG} to converge. Thus, we can
allow $\alpha$ to take on larger values to speed up the convergence
of Algorithm \ref{alg:mAPG} by means of a linear search procedure.
Specifically, let $L_{0}>0$, $\delta>0$ and $\rho\in(0,1)$. Then
the step size $\alpha$ in Step \ref{grady} of Algorithm \ref{alg:mAPG}
is set to $\alpha=L_{o}\rho^{m_{n}}$, where $m_{n}$ is the first
nonnegative smallest integer such that\begin{subequations} 
\begin{align}
\mathbf{z}^{n+1} & =P_{\mathcal{S}}\bigl(\mathbf{y}^{n}+L_{o}\rho^{m_{n}}\nabla f(\mathbf{y}^{n})\bigr),\\
f(\mathbf{z}^{n+1}) & \geq f(\mathbf{y}^{n})+\delta||\mathbf{z}^{n+1}-\mathbf{y}^{n}||^{2}.
\end{align}
\end{subequations}The above line search follows the Armijo
rule. The same line search procedure can be used for Step \ref{gradmu}
of Algorithm \ref{alg:mAPG}. We refer to these modifications as Algorithm
\ref{alg:mAPG} with line search. 

\section{Numerical Results}

In this section, we evaluate the performance of the proposed method
in terms of computational complexity and achieved spectral efficiency.
The users and the APs are uniformly dropped over a $D\times D$ \si{\km\squared}. The large-scale fading coefficient between the $m$-th AP and the
$k$-th user is generated as
\[
\beta_{mk}=\mathrm{PL}_{mk}.z_{mk},
\]
where $\mathrm{PL}_{mk}$ and $z_{mk}$ represent the path loss and
log-normal shadowing with mean zero and standard deviation $\sigma_{\textrm{sh}}$,
respectively. In this paper we adopt the three-slope path loss model
as in \cite{Ngo2018EE} in which $\mathrm{PL}_{mk}$ (in dB) is given
by 
\[\small
\mathrm{PL}_{mk}=\begin{cases}
-L-15\log_{10}(d_{1})-20\log_{10}(d_{0}) & d_{mk}<d_{0},\\
-L-35\log_{10}(d_{mk}) & d_{1}<d_{mk},\\
-L-15\log_{10}(d_{1})-20\log_{10}(d_{mk}) & \mathrm{otherwise},
\end{cases}
\]
where $L$ is a constant dependent on carrier frequency, $d_{mk}$ (in \si{\km}) is the distance between
the $m$-th AP and the $k$-th user, and $d_{0}$ and $d_{1}$ (both in
\si{\km}) are reference distances.%

Similar to \cite{Ngo2018EE}, we choose $L=140.7$ dB, $d_{0}=$
\SI{0.01}{\km} and $d_{1}=$ \SI{0.05}{\km}. We consider a system having
a bandwidth of $B=$ \SI{20}{\MHz}, the noise power density is $N_{0}=\text{\textminus}174$
(dBm/\si{\hertz}), and a noise figure of $9$ dB. The length of the
coherence time and the uplink training phase are set to $T_{p}=20$,
$T_{c}=200$, respectively. If not otherwise mentioned, we set the
power transmit power for downlink data transmission and uplink training
phase (before normalization) as $\zeta_{d}=1$
W and $\zeta_{p}=0.2$ W. Single antenna AP is considered in all numerical
experiments. These parameters are taken from \cite{Ngo2018EE}.

For comparison purpose, we modify the SCA-based method presented in
\cite{Ngo2018EE}, which is dedicated to the EEmax
problem. However, it can be easily modified to deal with the problems considered in this paper.
In this regard we use convex conic solver MOSEK \cite{MOSEKApS15}
through the modeling tool YALMIP \cite{YALMIP}.

In the first numerical experiment, we compare the convergence rate of the proposed method with the SCA-based method.
\begin{figure}[tbh]
	\centering
\subfigure[Spectral efficiency maximization]{\label{fig:convergence-1}\includegraphics[scale=0.7]{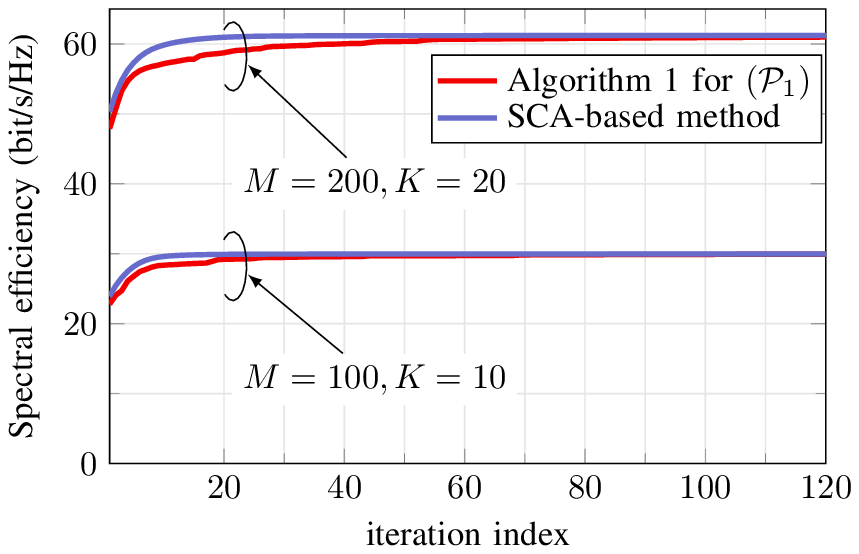}}\\
\subfigure[Minimum rate maximization]{\label{fig:convergence-2}\includegraphics[scale=0.7]{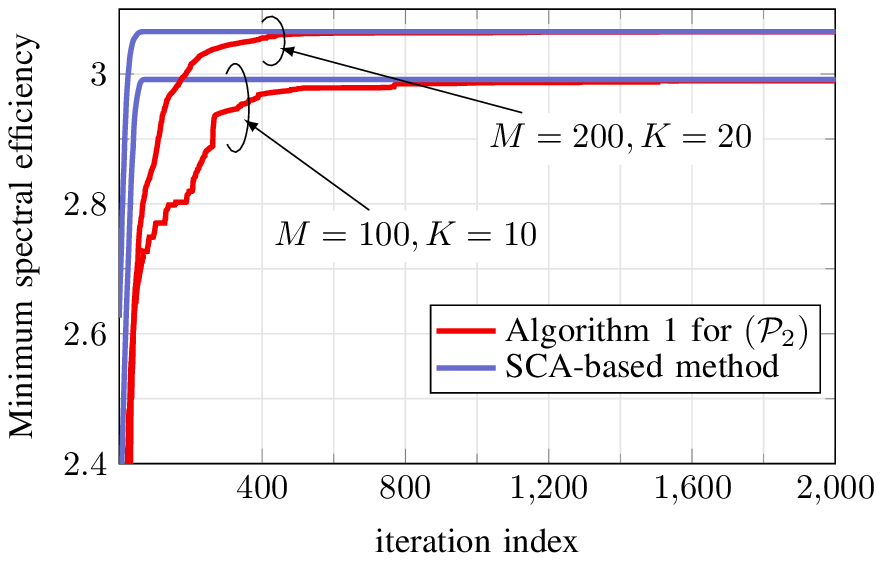}}
\caption{Total spectral efficiency and the minimum spectral efficiency versus
the number of iterations. The values of $M$ and $K$ are given explicitly
the figure. Each AP is equipped with one antenna.}
\label{fig:convergence}
\end{figure}
Figures~\ref{fig:convergence-1} and \ref{fig:convergence-2}
 show the convergence of the proposed method and the SCA-based method for the total spectral efficiency and the min-rate maximization problem, respectively.  We can see that  the proposed  and the SCA-based methods  achieve the same performance but the SCA-based method requires fewer iterations. However, the main advantage of our proposed method over the SCA-based method is that each iteration of the proposed method is very memory efficient, and hence, performs very fast. As a result, the run-time of the proposed method is far less than that of the SCA-based method as shown in Table~\ref{table: Table 1}. In Table~\ref{table: Table 1}, we report the actual run-time of both methods to solve the SEmax problem. Here, we execute our codes on a 64-bit Windows operating system with 16 GB RAM and Intel CORE i7, 3.7 GHz. Both iterative methods are terminated when the difference of the objective for the last $5$ iterations is less than $10^{-3}$. \vspace{-10pt}
\begin{table}[tbh]
\caption{Comparison of run-time (in seconds) between the proposed method and the SCA-based method. Here, $K=40$ and $D=1$.}
\label{table: Table 1}
\centering{}%
\begin{tabular}{c|c|c}
\hline 
APs & SCA Method & Proposed Method\tabularnewline
\hline 
200 & 330.84 & \textbf{2.88}\tabularnewline
\hline 
400 & 408.94 & \textbf{9.42}\tabularnewline
\hline 
800 & 1115.18 & \textbf{18.07}\tabularnewline
\hline 
1600 & 1648.09 & \textbf{49.45}\tabularnewline
\hline 
\end{tabular}
\end{table}

\begin{figure}[tbh]
\begin{centering}
{\includegraphics[scale=0.7]{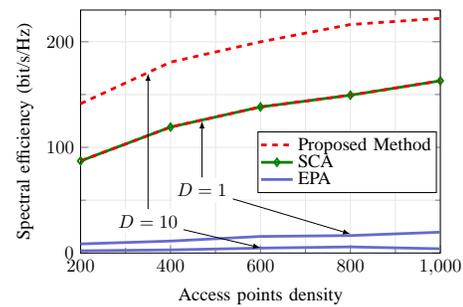}}
\par\end{centering}
\caption{Total spectral efficiency versus AP density. The number of users is
$K=40$.}
\label{fig:SE:APdensity}
\end{figure}

We next take advantage of the proposed method to explore the spectral efficiency performance of cell-free massive MIMO for a large metropolitan area. In particular, we investigate the performance for two cases $D=1$ and $D=10$. To obtain a fair comparison we keep the AP density,
defined as the number of APs per square kilometer, the same for both
cases. Note that the AP density of \num{1000} means \num{10000}
APs for the case of $D=10$, which has not been studied in the literature
previously. To appreciate the proposed method for this large-scale scenario, we compare it with the SCA method and the equal power allocation
(EPA) method where the power control coefficient $\eta_{mk}$ is given by, $\eta_{mk}=(\sum_{i=1}^{k}\nu_{mi})^{-1}$. The results in Figure~\ref{fig:SE:APdensity} are interesting. First, increasing the AP density improves the sum spectral efficiency of the system. Second, for the
same AP density, a larger area provides a better sum spectral efficiency. The reason is that for a larger area, the users that are served by the APs become far apart each other. As a result, the inter-user inference becomes
weaker, leading to an improved sum spectral efficiency. On the other hand, the EPA method
yields smaller spectral efficiency as the coverage area is larger because more power
should be to spent to the users having small path loss. The SCA method
produces the same spectral efficiency as the proposed APG method but it is unable to
run for $D=10$ on the system specifications mentioned above. Thus, the proposed scheme outclasses both  SCA-based  and EPA methods in
terms of improved sum spectral efficiency and larger coverage area.

\begin{figure}[tbh]
\begin{centering}
{\includegraphics[scale=0.7]{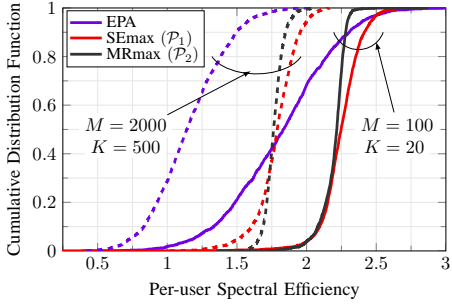}}
\par\end{centering}
\caption{CDF of per-user spectral efficiency for $(\mathcal{P}_{1})$ and $(\mathcal{P}_{2})$.
The parameters other than the mentioned are kept the same.}
\label{fig:CDFK20}
\end{figure}

Finally, we  compare the cumulative distribution function (CDF) of the per-user spectral efficiency for the EPA, SEmax, and  max-min fairness  power controls, see Figure~\ref{fig:CDFK20}. Here, we consider two scenarios:
(i) small-scale scenario: $M=100,K=20,T_{c}=200,T_{p}=20$, $D=1$; and (ii) large-scale
scenario: $M=2000,K=500,T_{c}=1000,T_{p}=200$, $D=1$.
As expected, the spread of CDF for the SEmax power control
is the larger than the max-min fairness power control.  On the other hand, the order is reversed in terms of fairness as the CDF curve of the max-min problem has the largest slope which shows that the spectral efficiencies of all the users are very close to each other. 
%,and thus,the fairness among the users is ensured. 
The per-user spectral efficiencies in the large-scale scenario are lesser than the ones in the small-scale scenario. This is because when the number  users increases, the inter-user interference increases which yields to the reduced performance.

\section{Conclusion}

We have considered the downlink of cell-free massive MIMO and aimed
to maximize the total and minimum spectral efficiencies, subject to a sum power constraint
at each AP. Conjugate beamforming has been adopted, resulting in a
power control problem for which an accelerated project gradient method
has been proposed. The proposed solution only requires the first order
information of the objective and, in particular, can be founded by
closed-form expressions. We have numerically shown that the proposed
method can achieve the same spectral efficiency as a known SCA-based method but with much lesser run-time. For the first time, we have evaluated the SE performance of cell-free massive MIMO for an area of \SI{10}{\km}$\times$\SI{10}{\km}, consisting of up to \num{10000}
APs. In this case, the achieved sum spectral efficiency can be up to 200 (bit/s/Hz).

\appendices{}%\section*{}
\section{Convergence Proof of Algorithm \ref{alg:mAPG} \label{sec:Convergence-Proof}}

The proof is due to \cite{Li2015}. We begin with by recalling an
important inequality of a $L$-smooth function. Specifically, for
a function $f(x)$ has the Lipschitz continuous gradient with a constant
$L$, the following inequality holds
\begin{equation}
f(\mathbf{y})\geq f(\mathbf{x})+\bigl\langle\nabla f\bigl(\mathbf{x}\bigr),\mathbf{y}-\mathbf{x}\bigr\rangle-\frac{L}{2}||\mathbf{y}-\mathbf{x}||^{2}.\label{eq:Lips:grad:inequality}
\end{equation}
The projection in Step \ref{gradmu} of Algorithm \ref{alg:mAPG}
can be written as
\begin{multline}
\mathbf{v}^{n+1}=\underset{\boldsymbol{\mu}\in\mathcal{S}}{\arg\min}\bigl\Vert\boldsymbol{\mu}-\boldsymbol{\mu}^{n}-\alpha\nabla f\bigl(\boldsymbol{\mu}^{n}\bigr)\bigr\Vert^{2}\\
=\underset{\boldsymbol{\mu}\in\mathcal{S}}{\arg\max}\ \bigl\langle\nabla f\bigl(\boldsymbol{\mu}^{n}\bigr),\boldsymbol{\mu}-\boldsymbol{\mu}^{n}\bigr\rangle-\frac{1}{2\alpha}||\boldsymbol{\mu}-\boldsymbol{\mu}^{n}||^{2}\label{eq:theta:project:rewrite},
\end{multline}
%where we have used the fact that $||\mathbf{a}-\mathbf{b}||^{2}=||\mathbf{a}||^{2}+||\mathbf{b}||^{2}+2\left\langle \mathbf{a},\mathbf{b}\right\rangle $.
%Note that when $\boldsymbol{\mu}=\boldsymbol{\mu}^{n}$, the objective
%in the above problem is $0$, 
and thus we have 
\begin{equation}
\bigl\langle\nabla f\bigl(\boldsymbol{\mu}^{n}\bigr),\mathbf{v}^{n+1}-\boldsymbol{\mu}^{n}\bigr\rangle-\frac{1}{2\alpha}||\mathbf{v}^{n+1}-\boldsymbol{\mu}^{n}||^{2}\geq0.
\end{equation}
Applying \eqref{eq:Lips:grad:inequality} yields
\begin{align}
f(\mathbf{v}^{n+1}) & \geq f\bigl(\boldsymbol{\mu}^{n}\bigr)+\bigl\langle\nabla f\bigl(\boldsymbol{\mu}^{n}\bigr),\mathbf{v}^{n+1}-\boldsymbol{\mu}^{n}\bigr\rangle\nonumber \\
 & -\frac{L}{2}\bigl\Vert\mathbf{v}^{n+1}-\boldsymbol{\mu}^{n}\bigr\Vert^{2}\nonumber \\
 & \geq f\bigl(\boldsymbol{\mu}^{n}\bigr)+\bigl(\frac{1}{2\alpha}-\frac{L}{2}\bigr)\bigl(\bigl\Vert\mathbf{v}^{n+1}-\boldsymbol{\mu}^{n}\bigr\Vert^{2}.\label{eq:theta:project:ineq}
\end{align}
It is easy to see that $f(\mathbf{v}^{n+1})\geq f\bigl(\boldsymbol{\mu}^{n}\bigr)$
if $\alpha<\frac{1}{L}$. From Step \ref{alg:mAPG:update}, if $f(\mathbf{z}^{n+1})\geq f(\mathbf{v}^{n+1})$,
then
\begin{align}
\boldsymbol{\mu}^{n+1} & =\mathbf{z}^{n+1},f\bigl(\boldsymbol{\mu}^{n+1}\bigr)=f(\mathbf{z}^{n+1})\geq f(\mathbf{v}^{n+1})\label{eq:extra:ineq}.
\end{align}
Similarly, if $f(\mathbf{z}^{n+1})<f(\mathbf{v}^{n+1})$, then 
\begin{align}
\boldsymbol{\mu}^{n+1} & =\mathbf{v}^{n+1},f\bigl(\boldsymbol{\mu}^{n+1}\bigr)=f(\mathbf{v}^{n+1}).\label{eq:theta:project:update}
\end{align}
From \eqref{eq:theta:project:ineq}, \eqref{eq:extra:ineq}, and \eqref{eq:theta:project:update}
we have 
\begin{equation}
f\bigl(\boldsymbol{\mu}^{n+1}\bigr)\geq f(\mathbf{v}^{n+1})\geq f\bigl(\boldsymbol{\mu}^{n}\bigr).
\end{equation}
Since the feasible set of the considered problems is compact convex,
the iterates $\{\mathbf{v}^{n}\}$ and $\{\boldsymbol{\mu}^{n}\}$
are both bounded and thus, $\{\boldsymbol{\mu}^{n}\}$ has accumulation
points. 
%As shown above, $f\bigl(\boldsymbol{\mu}^{n}\bigr)$ is nondecreasing,
%$f$ has the same value, denoted by $f^{\ast}$, at all the accumulation
%points. From \eqref{eq:theta:project:ineq} we have
%\begin{multline}
%f\bigl(\boldsymbol{\mu}^{n+1}\bigr)-f\bigl(\boldsymbol{\mu}^{n}\bigr)\geq f(\mathbf{v}^{n+1})-f\bigl(\boldsymbol{\mu}^{n}\bigr)\\
%\geq\bigl(\frac{1}{2\alpha}-\frac{L}{2}\bigr)\bigl(\bigl\Vert\mathbf{v}^{n+1}-\boldsymbol{\mu}^{n}\bigr\Vert^{2}
%\end{multline}
%which results in
%\begin{multline}
%\infty>f^{\ast}-f\bigl(\boldsymbol{\mu}^{1}\bigr)\geq\sum_{n=1}^{\infty}\bigl(\frac{1}{2\mu}-\frac{L}{2}\bigr)\bigl(\bigl\Vert\mathbf{v}^{n+1}-\boldsymbol{\mu}^{n}\bigr\Vert^{2}
%\end{multline}
%Since $\mu<\frac{1}{L}$ we can conclude that 
%\begin{equation}
%\bigl\Vert\mathbf{v}^{n+1}-\boldsymbol{\mu}^{n}\bigr\Vert\to0\ \textrm{as}\ n\to\infty\label{eq:iterate:converge}
%\end{equation}
%The convergence proof of Algorithm \ref{alg:mAPG} to 
We can prove that each accumulation
point of Algorithm \ref{alg:mAPG} is indeed a critical point
of $(\mathcal{P}_{1})$, following the same arguments as those in \cite{Li2015}
but skip the details due to the space limitation.

\section*{Acknowledgment}
This publication has emanated from research supported in part by a grant from
Science Foundation Ireland under grant number 17/CDA/4786.

\bibliographystyle{IEEEtran}
\bibliography{IEEEabrv,bibTex,references}

\end{document}